\documentclass[pre,superscriptaddress,showpacs,twocolumn,floatfix]{revtex4}
\usepackage{color,epsfig}
\usepackage{bm}
\usepackage{amsmath,amsfonts,amssymb,bm}
\usepackage{natbib}

\begin{document}

\title{Magnetically Driven Quantum Heat Engine}
\author{Enrique Mu\~{n}oz}
\affiliation{Facultad de F\'isica, Pontificia Universidad Cat\'olica de Chile, Vicu\~{n}a Mackenna 4860, Santiago, Chile.}
\author{Francisco ~J.~Pe\~na}
\affiliation{Instituto de F\'isica, Pontificia Universidad Cat\'olica de Valpara\'iso,
Av. Brasil 2950, Valpara\'iso, Chile.}
\date{\today}

\begin{abstract}
We studied the efficiency of two different schemes for a magnetically driven quantum heat engine, by considering as the "working substance"
a single non-relativistic particle trapped in a cylindrical potential well, in the presence
of an external magnetic field.
The first scheme is a cycle, composed of two adiabatic and two iso-energetic reversible trajectories in
configuration space. The trajectories are driven by a quasi-static modulation of the external magnetic field
intensity. The second scheme is a variant of the former, where the iso-energetic trajectories are
replaced by isothermal ones, along which the system is in contact with macroscopic thermostats. This second
scheme constitutes a quantum analogue of the classical Carnot cycle.
\end{abstract}


\maketitle

\section{Introduction}

In analogy with classical thermodynamics, a quantum heat engine generates useful mechanical work from heat, by means of a reversible sequence of transformations
(trajectories) in Hilbert's space, where the "working substance" is
of quantum mechanical nature
\cite{Bender_02,Bender_Brody_00,Wang_011,Wang_He_012,Quan_06,Arnaud_02,Latifah_011,Quan_Liu_07,
Scully_03,Scully_011,Quan_Zhang_05}. Several theoretical implementations
for a quantum heat engine
have been discussed in the literature, such as entangled states in a qubit \cite{Huang_013}, quantum mechanical versions of the Otto cycle \cite{Li_Negentropy_013,Lutz_014}, and
photocells \cite{Scully_03,Scully_011}. The photocell, where the working substance
are thermalized light quanta (photons), is at the same time inspired by and provides
a conceptual model for the mechanism of photosynthesis in plants and bacteria \cite{Dorfman_013}. In recent years, it has been proposed that if the reservoirs are also of quantum mechanical nature, these could be engineered
into quantum coherent states \cite{Scully_03,Scully_011} or into squeezed thermal states \cite{Lutz_014}, thus allowing for a theoretical enhancement of
the engine efficiency beyond the classical Carnot limit \cite{Lutz_014,Scully_011,Scully_03}.

One of the simplest theoretical implementations for a quantum heat engine is a system composed by a single-particle trapped in a one-dimensional potential well \cite{Wang_He_012,Wang_011,Bender_02,Bender_Brody_00,Latifah_011}.
The different trajectories can be driven by a quasi-static deformation
of the potential well, by applying an external force \cite{Bender_02,Bender_Brody_00,Munoz_Pena_012}.
Two different schemes of this process have been discussed in the literature, in the context
of a non-relativistic particle whose energy eigenstates are determined by
the Schr\"odinger spectrum \cite{Bender_02,Bender_Brody_00,Quan_Liu_07},
and more recently we studied an extension of the problem into the relativistic regime by considering the single-particle Dirac spectrum \cite{Munoz_Pena_012}.

In the present work, we propose yet a different alternative, by introducing
the notion of a magnetically driven quantum heat
engine.
 The basic idea is to combine the confining effects of a cylindrical potential well, which physically represents
an accurate model for a semiconductor quantum dot \cite{Jacak,Munoz_05}, and an external magnetic field. Assuming
that the intensity of the magnetic field can be modulated quasi-statically, then its confining effect over the
trapped particle will change accordingly, thus inducing transitions between the energy levels. These
single-particle states, as we show below, correspond to Landau levels\cite{Munoz_05} that combine the effect of both the geometric and magnetic confinements, as captured by the
effective frequency $\Omega = \sqrt{\omega_{d}^{2} + \omega_{B}^{2}/4}$.

\begin{figure}[tbp]
\centering
\epsfig{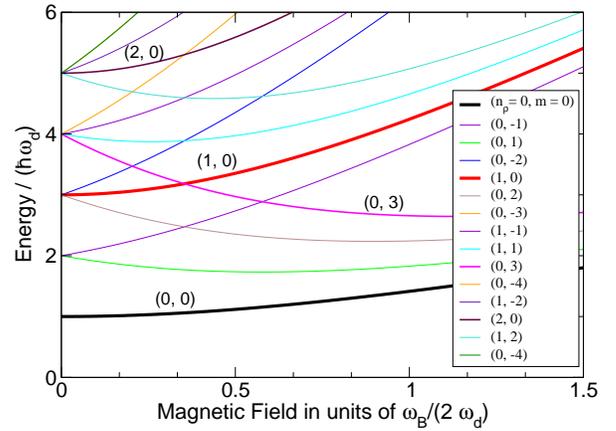}
\caption{(Color online) The single-particle energy spectrum for a cylindrical semiconductor quantum
dot under the presence of an external, uniform and static magnetic field, as defined in Eq.(\ref{eq6}). The horizontal axis
represents the magnitude of the field, in units of the ratio between the cyclotron frequency $\omega_{B} = e B/m^{*}$ and
the characteristic frequency of the dot $\omega_{d}$. Notice that
in general the energy levels $E_{n_{\rho},m}(B)$ are not degenerate, except for the discrete sequence of crossings that do not affect the ground state $(0,0)$. In particular, the lowest
excited state with $m = 0$, the level $(n_{\rho} = 1, m = 0)$, displays a single crossover with the level $(0, 3)$.}
\label{fig1}
\end{figure}

\section{The single-particle spectrum in a cylindrical well under external magnetic field}

Let us consider a single particle confined to a cylindrical potential well of the form
\begin{eqnarray}
V_{dot}(x,y) = \frac{m^{*}}{2}\omega_{d}^{2}\left(x^{2} + y^{2} \right).
\label{eq1}
\end{eqnarray}
Here, $m^{*}$ is the effective mass of the particle.
On top
of this "geometric" confinement, we shall assume a finite (constant) magnetic field along the z-axis:
\begin{equation}
{\mathbf{B}}=\hat{z} B.
\label{eq2}
\end{equation}
We shall adopt the symmetric gauge for the magnetic vector potential
\begin{equation}
{\mathbf{A}}=\frac{B}{2}(-y,x,0).
\label{eq3}
\end{equation}

Under the aforementioned considerations, the single-particle Hamiltonian representing this system is
\begin{equation}
\hat{H}=\frac{1}{2m^*}\left[\left(p_x-\frac{e\,By}{2}\right)^2+\left(p_y+\frac{e\,Bx}{2}\right)^2\right]+V_{dot}(x,y).
\label{eq4}
\end{equation}
The single-particle eigenstates are obtained as solutions of the eigenvalue equation:
\begin{equation}
\hat{H}|\psi\rangle=E|\psi\rangle,
\label{eq5}
\end{equation}
and they correspond to Landau levels \cite{Munoz_05} with the energy spectrum:
\begin{equation}
E_{n_\rho,m}(B)=\hbar \Omega(2n_\rho+\mid{m}\mid+1)-m\frac{\hbar\omega_B}{2}.
\label{eq6}
\end{equation}
Here, $n_\rho=0,1,2,...$ and $m=0,\pm1,\pm2,..$
are the radial and azimuthal quantum numbers, respectively.
We have defined the effective frequency
\begin{equation}
\Omega=\sqrt{\omega_d^2+\frac{\omega_B^2}{4}},
\label{eq7}
\end{equation}
where
\begin{equation}
\omega_B = \frac{eB}{m^*}
\label{eq8}
\end{equation}
is the standard definition for the cyclotron frequency. The eigenfunctions are expressed in terms of
associated Laguerre polynomials \cite{Munoz_05}
\begin{eqnarray}
\langle {\mathbf{r}} | \psi_{\mathbf{n}}(B)\rangle &=& \sqrt{\frac{2}{l_{e,B}^{2}}}
\sqrt{\frac{n_{\rho}!}{\left(n_{\rho} + |m| \right)!}}\left(\frac{\rho}{l_{e,B}} \right)^{|m|}\nonumber\\
&&\times e^{-\frac{\rho^{2}}{2 l_{e,B}^{2}}}L_{n_{\rho}}^{|m|}\left(\frac{\rho^{2}}{l_{e,B}^{2}} \right),
\label{eq9}
\end{eqnarray}
where $l_{e,B} = \sqrt{\hbar/\left(m^{*}\Omega \right)}$ is the effective Landau radius
that characterizes the combined confining effects of the potential as well as the external magnetic field, and
$\mathbf{n}\equiv (n_{\rho},m)$.
The single-particle spectrum defined by Eq.(\ref{eq6}) is depicted in Fig.~\ref{fig1}, in units
of the quantum dot characteristic energy $\hbar \omega_{d}$, as a function of the
external magnetic field, for the
first fourteen eigenstates. From the figure it is clear that, even  at arbitrary large magnetic fields, the ground state $(n_{\rho}=0, m=0)$ is non-degenerate. The excited states, on the other hand, are
non-degenerate except for a discrete set of crossovers as the field intensity increases. In particular, the excited state
$(n_{\rho} = 1, m = 0)$ exhibits only a single crossover with
the state $(n_{\rho} = 0, m = 3)$, a feature to be discussed
later on in the context of the Iso-energetic cycle.

A cylindrical potential well, like the one we consider here, is a standard approximation
for the effective confinement in semiconductor quantum dots \cite{Jacak,Munoz_05}.
For instance, in a cylindrical GaAs quantum dot, a typical value for the effective mass would be $m^{*} \sim 0.067 m_{0}$ \cite{Jacak,Munoz_05}, with a typical radius $l_{d}\sim 20 - 100$ nm \cite{Jacak}. In our present analysis, for simplicity we do not include the Zeeman term, which is typically negligible for GaAs semiconductor nanostructures even at high magnetic field intensities \cite{Jacak}.

\section{A single-particle Quantum heat Engine}

As the "working substance" for a quantum heat engine, let us consider a statistical ensemble
of replicas of a single-particle system, where each replica may be in any of the different eigenstates
of the Hamiltonian Eq.(\ref{eq4}).
The single-particle system
is then in a statistically mixed quantum state\cite{vonNeumann}, described by the density
matrix operator $\hat{\rho} = \sum_{\mathbf{n}}p_{\mathbf{n}}(B)|\psi_{\mathbf{n}}(B)\rangle\langle\psi_{\mathbf{n}}(B)|$, with $|\psi_{\mathbf{n}}(B)\rangle$
an eigenstate of the single-particle Hamiltonian Eq.(\ref{eq4}) for a given (fixed) magnetic field intensity $B$. To alleviate the notation, we introduced the two-valued index $\mathbf{n}\equiv (n_{\rho},m)$
to enumerate the eigenstates of the Hamiltonian defined in Eq.(\ref{eq6}). This density matrix operator
is stationary, since in the absence of an external perturbation\cite{vonNeumann} $i\hbar\partial_{t}\hat{\rho} = [\hat{H},\hat{\rho}] = 0$. Here, the coefficient $0 \le p_{\mathbf{n}}(B)\le 1$
represents the probability for the system, within the statistical ensemble, to be in the particular state
$|\psi_{\mathbf{n}}(B)\rangle$. Therefore, the $\{p_{\mathbf{n}}(B)\}$ satisfy the normalization condition
\begin{eqnarray}
{\rm{Tr}}\hat{\rho} = \sum_{\mathbf{n}}p_{\mathbf{n}}(B) = 1.
\label{eq10}
\end{eqnarray}

In the context of Quantum Statistical Mechanics, entropy is defined according
to von Neumann \cite{vonNeumann,Tolman} as $S = -k_{B}{\rm{Tr}}\hat{\rho}\ln\hat{\rho}$.
Since in the energy eigenbasis the equilibrium density matrix operator is diagonal, the entropy reduces to the explicit expression
\begin{eqnarray}
S(B) = -k_{B}\sum_{\mathbf{n}}p_{\mathbf{n}}(B)\ln\left(p_{\mathbf{n}}(B)\right).
\label{eq11}
\end{eqnarray}
In our notation, we emphasize the explicit dependence of the energy eigenstates $\{|\psi_{\mathbf{n}}(B)\rangle\}$, as well as the probability coefficients $\{p_{\mathbf{n}}(B)\}$,
on
the intensity of the external magnetic field $B$.

The ensemble-average energy of the quantum single-particle system is
\begin{eqnarray}
E \equiv \langle \hat{H} \rangle = {\rm{Tr}}(\hat{\rho}\hat{H}) = \sum_{\mathbf{n}}p_{\mathbf{n}}(B) E_{\mathbf{n}}(B).
\label{eq12}
\end{eqnarray}

For the statistical ensemble just defined, we consider two different schemes
for a quantum analogue of a thermodynamic heat engine. The first one, that we shall refer to as the Iso-energetic cycle,
consists on four stages of reversible trajectories: two iso-entropic
and two iso-energetic ones, as originally proposed
by Bender et al.~\cite{Bender_02,Bender_Brody_00} in the context of a Schr\"odinger particle,
and more recently extended by us to a relativistic Dirac particle \cite{Munoz_Pena_012}.
Along the iso-energetic trajectories, the ensemble-average
energy Eq.(\ref{eq12})
is conserved, while during the iso-entropic
ones, the von Neumann entropy defined by Eq.(\ref{eq11}) remains constant. We distinguish
this first scheme from the quantum Carnot cycle to be discussed next, where
the iso-energetic trajectories in Hilbert's space are substituted by
isothermal processes, as the system is brought into thermal equilibrium with
macroscopic reservoirs at temperatures $T_{C} \le T_{H}$, respectively.
\begin{figure}[tbp]
\centering
\epsfig{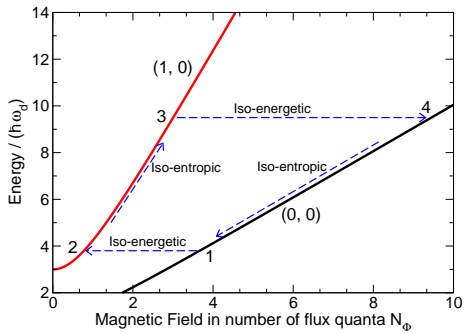}
\caption{(Color online)The Iso-energetic cycle for the effective two-level system composed by the ground state $(n_{\rho} = 0, m = 0)$, and the
first accessible excited state $(n_{\rho} = 1, m = 0)$ compatible with the selection rule $\mathcal{T}_{1\rightarrow 2} \propto \delta_{m_{1}, m_{2}}$, as explained in the main text. The cycle involves two iso-energetic trajectories and two iso-entropic ones.
\label{fig2}
}
\end{figure}

\section{The Iso-energetic Cycle}

The Iso-energetic cycle, a scheme for a quantum heat-engine originally proposed by Bender et al.~\cite{Bender_02,Bender_Brody_00} in the context of a single Schr\"odinger particle, and more recently
extended by us to a relativistic Dirac particle \cite{Munoz_Pena_012},
is composed by two iso-entropic and two iso-energetic trajectories.
In particular, during the
iso-energetic trajectories, the "working substance" must exchange energy
with an energy reservoir \cite{Wang_011,Wang_He_012}. A possible practical realization of
this cycle was proposed in Ref.~\cite{Wang_011}, where the working substance
exchanges energy with single-mode radiation in a cavity, which acts as an energy reservoir.

The system trajectories in Hilbert's space are assumed to be driven by reversible quasi-static processes,
in which the walls of the
potential well are deformed quasi-statically by an applied external force, such that the distance $L$ is modified
accordingly \cite{Bender_02,Bender_Brody_00,Munoz_Pena_012}.
In this work, we propose a similar concept (see Fig.~\ref{fig2}), but the confinement length to be modified is
the effective Landau radius introduced in Eq.(\ref{eq9})
\begin{eqnarray}
l_{e,B}= \sqrt{\hbar/ (m^{*}\Omega)} = \left(l_{d}^{-4} + l_{B}^{-4}/4 \right)^{-1/4},
\label{eq13}
\end{eqnarray}
with $l_{d} = \sqrt{\hbar/(m_{*}\omega_{d})}$ representing the
geometric confinement imposed by the cylindrical potential well, while
\begin{eqnarray}
l_{B} = \sqrt{\hbar/ (m^{*}\omega_{B})}
\label{eq14}
\end{eqnarray}
is the confinement scale imposed by the magnetic field. The effective Landau radius $l_{e,B}$
can be modified through $l_{B}$, by quasi-statically adjusting the intensity of the external magnetic field. Along these trajectories,
the total change in the ensemble average energy of the system is given by
\begin{eqnarray}
d E &=& \sum_{\mathbf{n}}p_{\mathbf{n}}(B) dE_{\mathbf{n}}(B) + \sum_{n} dp_{\mathbf{n}}(B) E_{\mathbf{n}}(B)\nonumber\\
&=&\left(\delta  E \right)_{\{p_{\mathbf{n}}(B)\} = cnt.}
+ \left(\delta  E  \right)_{\{E_{\mathbf{n}}(B)\} = cnt.},
\label{eq15}
\end{eqnarray}
where we have introduced the two-valued index $\mathbf{n}\equiv (n_{\rho},m)$
to enumerate the eigenstates of the Hamiltonian, defined in Eq.(\ref{eq6}). The first term in Eq.(\ref{eq15}) represents the total energy change due to an iso-entropic process, whereas
the second term represents a trajectory where the energy spectrum remains rigid.

\begin{figure}[tbp]
\centering
\epsfig{file=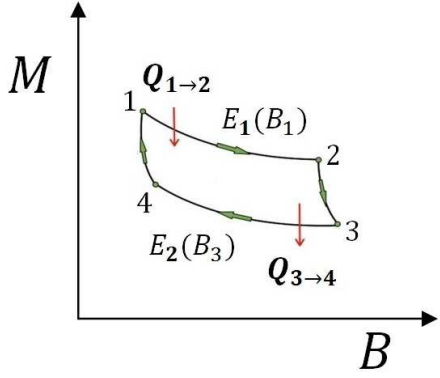,width=0.7\columnwidth,clip=}
\caption{(Color online)Pictorial description of the Magnetization versus external magnetic field for the idealized Iso-energetic cycle.
\label{fig3}
}
\end{figure}

Let us first consider an iso-entropic process, defined by the constraint \cite{note1} $\{p_{\mathbf{n}}(B)\} = cnt$.
Under quasi-static conditions, the work performed in varying the external magnetic field $B$
is related to the magnetization $M = -\left(\partial E/\partial B\right)_{S}$ of the system (see Fig.\ref{fig3}), $dW = -M dB$.
Therefore, when
the magnetic field changes from $B=B_{\gamma}$ to $B=B_{\delta}$, the total work performed by the system is
\begin{eqnarray}
W_{\gamma \rightarrow \delta} &=& \int_{B_{\gamma}}^{B_{\delta}}dB \left(\frac{\partial E}{\partial B}\right)_{\{p_{\mathbf{n}}(B_{\gamma})=p_{\mathbf{n}}(B_{\delta})\}= cnt.}\nonumber\\
&=& \sum_{\mathbf{n}}p_{\mathbf{n}}(B_{\gamma}) \left[E_{\mathbf{n}}(B_{\delta}) - E_{\mathbf{n}}(B_{\gamma})\right].
\label{eq16}
\end{eqnarray}

Notice that our sign convention is such that,
for an expansion process $l_{e,B_{\delta}} > l_{e,B_{\gamma}}$, the work performed by the
system is negative\cite{Callen}, indicating that the ensemble-averaged energy is decreasing during expansion.

Let us now consider an iso-energetic process, that is, a trajectory in Hilbert space defined
by the equation
$d E = 0$. The solution to this equation, for $B \in [B_{\delta},B_{\gamma}]$, is given by the path
\begin{eqnarray}
\sum_{\mathbf{n}} p_{\mathbf{n}}(B)E_{\mathbf{n}}(B)  = \sum_{\mathbf{n}} p_{\mathbf{n}}(B_{\gamma})E_{\mathbf{n}}(B_{\gamma}),
\label{eq17}
\end{eqnarray}
along with the normalization condition Eq.(\ref{eq10}).
Clearly, by definition an iso-energetic process satisfies
\begin{eqnarray}
d E  =  \delta W_{\gamma\rightarrow \delta} + \delta Q_{\gamma\rightarrow \delta} = 0,
\label{eq18}
\end{eqnarray}
with $\delta W_{\gamma\rightarrow \delta} \equiv \left(\delta  E  \right)_{\{p_{\mathbf{n}}(B)\} = cnt.}$ and
$\delta Q_{\gamma\rightarrow \delta} \equiv \left(\delta  E  \right)_{\{E_{\mathbf{n}}(B)\} = cnt.}$.
The integral of Eq.(\ref{eq18}) along the trajectory $B_{\gamma}\rightarrow B_{\delta}$ yields
\begin{eqnarray}
\Delta  E = W_{\gamma\rightarrow \delta} + Q_{\gamma\rightarrow \delta} = 0.
\label{eq19}
\end{eqnarray}
Here, the first term $W_{\gamma\rightarrow \delta}$ corresponds to the magnetic work performed by the system when changing its induced magnetization due to the applied field, at constant
total energy. The second term $Q_{\gamma\rightarrow \delta} = -W_{\gamma\rightarrow \delta}$ corresponds
to the amount of energy exchanged by the system
with the environment, in order to rearrange its internal level occupation. According to the previous analysis, the heat exchanged by the system with the environment along the iso-energetic process is given by
\begin{eqnarray}
Q_{\gamma\rightarrow \delta} = \sum_{\mathbf{n}}\int_{B_{\gamma}}^{B_{\delta}} E_{\mathbf{n}}(B)\frac{dp_{{\mathbf{n}}}(B)}{dB}\,dB.
\label{eq20}
\end{eqnarray}
Evidently, Eq.(\ref{eq17}) combined with the normalization condition Eq.(\ref{eq10}) are not enough to uniquely define the coefficients $p_{\mathbf{n}}(B)$
along an iso-energetic trajectory. An exception is the case
when the energy scale of all the processes involved is such that
only transitions between two adjacent levels are possible. Since the only driving force to induce transitions is a quasi-static variation of the magnetic field intensity, the azimuthal symmetry of the Hamiltonian Eq.(\ref{eq4}) is conserved at any moment along this process. Therefore, the transition probability amplitudes impose the selection rule ${\mathcal{T}}_{{\mathbf{1}}\rightarrow{\mathbf{2}}} \propto \int d\varphi e^{i(m_{1} - m_{2})\varphi} = \delta_{m_{1},m_{2}}$, and hence the azimuthal quantum numbers of initial and final states must be the same $m_{1}=m_{2}$, that is
angular momentum $L_{z}$ is conserved for such a transition.

Let us focus on the
particular case when the two states involved are the ground state and the first accessible excited state, respectively: $\mathbf{1} \equiv (n_{\rho} = 0, m = 0)$ and $\mathbf{2} \equiv (n_{\rho} = 1, m = 0)$ (see Fig.~\ref{fig1} and Fig.~\ref{fig2}). It is clear that transitions between these two energy levels, when the quasi-static
variation of the magnetic field intensity is the only driving force, are allowed
by symmetry, since $m_1 = m_2 = 0$ for both eigenstates. Moreover, by looking at
Fig.~\ref{fig1} it is clear that $(n_{\rho}=1, m=0)$ is the lowest excited state
accessible from the ground state $(n_{\rho}=0, m=0)$ that respects the selection rule
imposed by angular momentum conservation. From Fig.~\ref{fig1} it is also evident that
a crossover occurs between the excited states $(n_{\rho}=1, m=0)$ and $(n_{\rho} = 0, m=3)$. However, transitions between the ground state $(n_{\rho} = 0, m = 0)$ and the excited state $(n_{\rho} = 0, m = 3)$, or between $(n_{\rho} = 0, m = 3)$ and $(n_{\rho} = 1, m = 0)$ are forbidden by angular momentum conservation, if the quasi-static
variation of the magnetic field intensity is the only driving force, as discussed above. The next excited state that would respect the symmetry is $(n_{\rho} = 2, m = 0)$ but, as clearly seen in Fig.~\ref{fig1}, this one is quite high in energy
and does not cross at any point with $(0, 0)$ nor $(1, 0)$.
Therefore, under these considerations, the ground state $(n_{\rho} = 0, m = 0)$ and the excited state $(n_{\rho}=1, m=0)$ constitute an effective two-level system.

In practice, single quantum dots can be prepared in the ground state $(0, 0)$ by coupling them to a modified high-Q single-defect cavity \cite{Englund_05}. It has
been shown experimentally that a high-Q cavity can enhance the spontaneous emission rate
in single quantum dots up to a factor of 8 \cite{Englund_05}, due to Purcell's effect \cite{Agarwal}. Therefore, after a very short time the dot will decay
towards the ground state regardless of its initial condition. Once the system
has been prepared in its ground state, the iso-energetic cycle can start
by quasi-statically tuning the static magnetic field in the absence of external
radiation sources.

\begin{figure}[tbp]
\centering
\epsfig{file=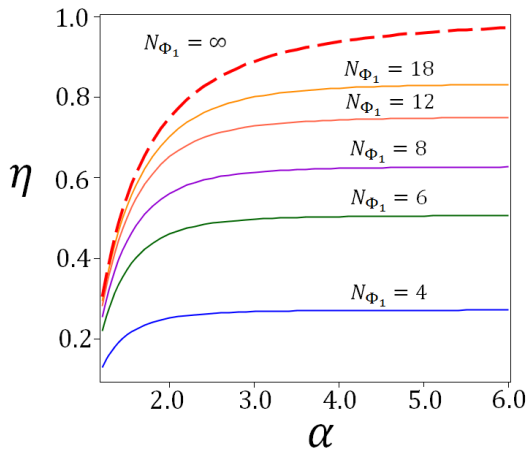,width=0.9\columnwidth,clip=}
\caption{(Color online) The efficiency of the Iso-energetic cycle, calculated from Eq.(\ref{eq34}),
is represented as a function of the expansion parameter $\alpha > 1$. Different values of the initial magnetic field
in the cycle $B_{1}$, expressed in terms of the number of flux quanta $N_{\Phi_{1}}$, are
compared. We find that the asymptotic limit represented by Eq.(\ref{eq35}) (red dashed line in the figure) is achieved in practice for $N_{\Phi_{1}}>30$.}
\label{fig4}
\end{figure}

 The effective two-level system constituted by the states $(0, 0)$ and $(1, 0)$ as described before, along with the different trajectories involved in
 the cycle, is represented in Fig.~\ref{fig4}.
Combining Eq.(\ref{eq17}) with the normalization condition Eq.(\ref{eq10}), the iso-energetic trajectories are described by
\begin{eqnarray}
p_{\mathbf{1}}(B) &=& \frac{E_{\mathbf{2}}(B_{1}) - E_{\mathbf{2}}(B)}{E_{\mathbf{1}}(B) - E_{\mathbf{2}}(B)} + \frac{E_{\mathbf{1}}(B_{1}) - E_{\mathbf{2}}(B_{1})}{E_{\mathbf{1}}(B) - E_{\mathbf{2}}(B)}p_{\mathbf{1}}(B_{1}),\nonumber\\
\label{eq21}
\end{eqnarray}
with $p_{\mathbf{2}}(B) = 1- p_{\mathbf{1}}(B)$ after the normalization condition Eq.(\ref{eq10}).
The heat exchanged by the system with the environment during the iso-energetic trajectory connecting the initial and final states $B_{1}\rightarrow B_{2}$,
is given by the expression
\begin{eqnarray}
Q_{1\rightarrow 2} &=& \left[E_{\mathbf{2}}(B_{1}) + \left(E_{\mathbf{1}}(B_{1})
- E_{\mathbf{2}}(B_{1})\right)p_{\mathbf{1}}(B_{1})\right]\nonumber\\
&&\times\ln\left[\frac{E_{\mathbf{1}}(B_{2})
-E_{\mathbf{2}}(B_{2})}{E_{\mathbf{1}}(B_{1})-E_{\mathbf{2}}(B_{1})}\right],
\label{eq22}
\end{eqnarray}
where the spectrum $E_{\mathbf{n}}(B)$ was defined in Eq.(\ref{eq6}).

For the effective two-level system described in Fig.~\ref{fig2}, we consider the cycle which starts in the ground state with $p_{\mathbf{1}}(B_{1}) = 1$. Then, the system experiences an iso-energetic expansion from $l_{B_{1}}\rightarrow l_{B_{2}}> l_{B_{1}}$, followed by an iso-entropic expansion from $l_{B_{2}}\rightarrow l_{B_{3}}>l_{B_{2}}$. Then, it experiences an iso-energetic compression $l_{B_{3}}\rightarrow l_{B_{4}}< l_{B_{3}}$, to finally return to its initial ground state through an iso-entropic compression $l_{B_{4}}\rightarrow l_{B_{1}}$.

We shall assume that
the final state after the iso-energetic process $l_{B_{1}}\rightarrow l_{B_{2}}$ corresponds to maximal expansion, that is, the system ends completely localized in the excited state $\mathbf{n}=\mathbf{2}$. In this condition, we have
\begin{eqnarray}
p_{\mathbf{1}}(B_{2}) = 0,\,\,\,\,p_{\mathbf{2}}(B_{2}) = 1.
\label{eq23}
\end{eqnarray}
The conservation of total energy between the initial and final states connected through the iso-energetic process leads to the equation
\begin{eqnarray}
p_{\mathbf{1}}(B_{1})E_{\mathbf{1}}(B_{1}) = p_{\mathbf{2}}(B_{2})E_{\mathbf{2}}(B_{2}),
\label{eq24}
\end{eqnarray}
where $p_{\mathbf{1}}(B_{1}) = p_{\mathbf{2}}(B_{2}) = 1$ for maximal expansion. Therefore,
given the spectrum in Eq.(\ref{eq6}), Eq.(\ref{eq24})
implies that $l_{B_{2}}/l_{B_{1}} =  \alpha_{1}$, where $\alpha_{1}$ is determined by the condition
\begin{eqnarray}
\hbar\omega_{d}\sqrt{1 + N_{\Phi_{1}}^{2}} = 3\hbar\omega_{d}\sqrt{1 + \frac{N_{\Phi_{1}}^{2}}{\alpha_{1}^{4}}}.
\label{eq25}
\end{eqnarray}
Here, we have defined $N_{\Phi_{1}} = l_{d}^{2}/(2 l_{B_{1}}^{2}) = \Phi_{B_{1}}/\Phi_{0}$ as the
number of flux quanta $\Phi_{0} = h/(2e)$ piercing the area $\pi l_{d}^{2}$.
Equation (\ref{eq25}) possesses physically
meaningful solutions
\begin{eqnarray}
\alpha_{1} = \frac{\sqrt{3N_{\Phi_{1}}}}{\left(N_{\Phi_{1}}^{2} - 8 \right)^{1/4}}
\label{eq26}
\end{eqnarray}
when $N_{\Phi_{1}} > 2\sqrt{2}$. Therefore, the minimal
initial value of the external field required to perform the cycle is $B_{1,min} = 4\sqrt{2}\hbar/(e l_{d}^{2})$. For instance, if one considers a typical size
of a semiconductor quantum dot of $l_{d} = 70$ nm \cite{Jacak}, the minimal initial field is $B_{1,min} \sim 4.8$ T. The heat exchanged with the environment along this first stage of the cycle is calculated after Eq.(\ref{eq22})
\begin{eqnarray}
Q_{1\rightarrow 2}= E_{\mathbf{1}}(B_{1})\ln\left[\frac{E_{\mathbf{2}}(B_{1})
- E_{\mathbf{1}}(B_{1})}{E_{\mathbf{2}}(\alpha_{1}^{-2}B_{1})-E_{\mathbf{1}}(\alpha_{1}^{-2}B_{1})}\right],
\label{eq27}
\end{eqnarray}
while the work performed is $W_{1\rightarrow 2} = -Q_{1\rightarrow 2}$.

The next process along the cycle is an iso-entropic expansion (see Fig.\ref{fig2}), characterized by the
condition $p_{\mathbf{2}}(B_{2}) = p_{\mathbf{2}}(B_{3}) = 1$. We shall define
the expansion parameter $\alpha \equiv l_{B_{3}}/l_{B_{2}} > 1$. Notice that $\alpha > 1$ can be arbitrarily chosen. The work
performed during this stage, with $l_{B_{2}} = \alpha_{1} l_{B_{1}}$ is
calculated from Eq.(\ref{eq16})
\begin{eqnarray}
W_{2\rightarrow 3} &=& 3\hbar\omega_{d}\left(\sqrt{1 + \frac{N_{\Phi_{1}}^{2}}{(\alpha_{1}\alpha)^{4}}} - \sqrt{1 + \frac{N_{\Phi_{1}}^{2}}{\alpha^{4}}} \right).
\label{eq28}
\end{eqnarray}
The cycle continues with a maximal compression process from $l_{B_{3}} = \alpha_{1} \alpha l_{B{1}}$
to $l_{B_{4}} = \alpha_{3}\alpha_{1}\alpha l_{B{1}}$ under iso-energetic conditions (see Fig.\ref{fig2}). The condition for energy conservation is in this case
similar to Eq.(\ref{eq24}), implying $p_{\mathbf{2}}(B_{3}) = p_{\mathbf{1}}(B_{4}) = 1$ and
\begin{eqnarray}
\hbar\omega_{d}\sqrt{1 + \frac{N_{\Phi_{1}}^{2}}{\left(\alpha_{1}\alpha_{3}\alpha \right)^{4}}} = 3\hbar\omega_{d}\sqrt{1 + \frac{N_{\Phi_{1}}^{2}}{\left(\alpha_{1}\alpha\right)^{4}}}.
\label{eq29}
\end{eqnarray}
The solution to Eq.(\ref{eq29}) fixes the value for the compression
coefficient $\alpha_{3}<1$
\begin{eqnarray}
\alpha_{3} = \frac{N_{\Phi_{1}}^{1/2}}{\left(8\left(\alpha_{1}\alpha \right)^{4} + 9 N_{\Phi_{1}}^{2} \right)^{1/4}}.
\label{eq30}
\end{eqnarray}

The heat exchanged
by the system with the environment along this process, applying Eq.(\ref{eq20}), is given by the expression
\begin{eqnarray}
Q_{3\rightarrow 4}&=& E_{\mathbf{2}}\left(\frac{B_{1}}{(\alpha\alpha_{1})^{2}}\right)\nonumber\\
&&\times\ln\left[\frac{E_{\mathbf{1}}(\frac{B_{1}}{\left(\alpha\alpha_{1}\alpha_{3}\right)^{2}})
- E_{\mathbf{2}}(\frac{B_{1}}{\left(\alpha\alpha_{1}\alpha_{3}\right)^{2}})}{E_{\mathbf{1}}(\frac{B_{1}}{(\alpha\alpha_{1})^{2}})-E_{\mathbf{2}}(\frac{B_{1}}{(\alpha\alpha_{1})^{2}})}\right]
\label{eq31}
\end{eqnarray}
and the work performed is $W_{3\rightarrow 4} = - Q_{3\rightarrow 4}$.
The last path along the cycle is an adiabatic process (see Fig.\ref{fig2}), which returns the system to its initial
ground state with $p_{\mathbf{1}}(B_{4}) = p_{\mathbf{1}}(B_{1}) = 1$. The work performed during
this final stage, as obtained by applying Eq.(\ref{eq16}), is given by
\begin{eqnarray}
W_{4\rightarrow 1} = \hbar\omega_{d}
\left(\sqrt{1 + N_{\Phi_{1}}^{2}} - \sqrt{1 + \frac{N_{\Phi_{1}}^{2}}{\left(\alpha\alpha_{1}\alpha_{3} \right)^{4}}} \right)
\label{eq32}
\end{eqnarray}
It follows from Eqs.(\ref{eq28}),(\ref{eq32}), in combination with Eqs.(\ref{eq25}),(\ref{eq29})
that the net work along the iso-entropic trajectories cancels, $W_{2\rightarrow 3} + W_{4\rightarrow 1} = 0$.

The efficiency of the cycle is thus defined by the ratio
\begin{eqnarray}
\eta = 1 - \left|\frac{Q_{3\rightarrow 4}}{Q_{1\rightarrow 2}}\right|.
\label{eq33}
\end{eqnarray}
When substituting the corresponding expressions from Eq.(\ref{eq27}) and Eq.(\ref{eq31}) into Eq.(\ref{eq33}),
we obtain the explicit analytical result
\begin{eqnarray}
\eta\left(N_{\Phi_{1}},\alpha \right) = 1 - 3\frac{\Theta_{1}\left(\alpha\alpha_{1} \right)}{\Theta_{1}(1)}
\frac{ln\left[\frac{\Theta_{1}\left(\alpha\alpha_{1}\alpha_{3} \right)}{\Theta_{1}\left(\alpha\alpha_{1}\right)} \right]}{ln\left[\frac{\Theta_{1}(1)}{\Theta_{1}\left(\alpha_{1} \right)}\right]},
\label{eq34}
\end{eqnarray}
where we have defined $\Theta_{1}\left(\alpha \right) = \sqrt{1 + N_{\Phi_{1}}^{2}/\alpha^{4}}$.
The trend of the efficiency is shown in Fig.~\ref{fig4} as a function of the expansion parameter $\alpha$,
for different values of the initial external field $B_{1}$ (expressed in terms of $N_{\Phi_{1}}$). For very large fields $N_{\Phi_{1}}\gg 1$, one has
from Eq.(\ref{eq26}) and Eq.(\ref{eq30}) that $\alpha_{1} = 1/\alpha_{3} = \sqrt{3}$, and hence the efficiency tends to the asymptotic limit
\begin{eqnarray}
\eta \rightarrow 1 - 1/\alpha^{2},\,\,\,N_{\Phi_{1}}\gg 1.
\label{eq35}
\end{eqnarray}
Remarkably, this asymptotic result has been obtained before for an Iso-energetic cycle
driven by a mechanical external force, both in the Schr\"odinger \cite{Bender_02} as well
as in the low-energy limit for the Dirac particle case \cite{Munoz_Pena_012}.
This suggests that it may represent a
universal maximal efficiency for any quantum mechanical engine based on the Iso-energetic cycle construction, just as the Carnot efficiency is to classical macroscopic heat engines.

\section{The Quantum Carnot cycle}

\begin{figure}[tbp]
\centering
\epsfig{file=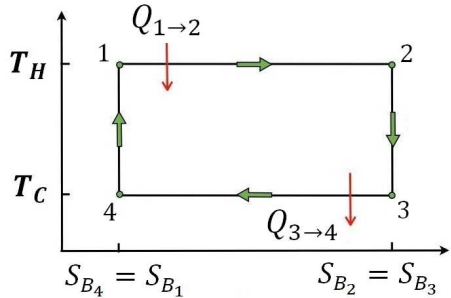,width=0.7\columnwidth,clip=}
\caption{(Color online) The quantum Carnot cycle discussed in this section is pictorially
represented. The isothermal
trajectories are achieved by bringing the system into contact with
macroscopic thermal reservoirs at temperatures $T_{H}> T_{C}$, respectively.}
\label{fig5}
\end{figure}

In this section, we shall discuss a quantum mechanical version of the Carnot cycle,
as applied to the statistical ensemble of single-particle systems under consideration.
The thermodynamic cycle which
defines the corresponding heat engine is composed of four stages or trajectories
in Hilbert's space: Two isothermal and two iso-entropic
processes, as depicted in Fig.~\ref{fig5}.

Along the first stage of the cycle, the system is brought into contact with a thermal reservoir
at temperature $T_{H}$. By keeping isothermal conditions, the Landau radius is expanded from $l_{e,B_{1}}\rightarrow l_{e,B_{2}}$. Since thermal equilibrium
with the reservoir is assumed along this process, the von Neumann entropy of the system
achieves a maximum for the Boltzmann distribution \cite{vonNeumann,Tolman}
\begin{eqnarray}
p_{\mathbf{n}}(B,\beta_{H}) = \left[Z(B,\beta_{H})\right]^{-1}e^{-\beta_{H}E_{\mathbf{n}}(B)},
\label{eq36}
\end{eqnarray}
with $\beta = (k_{B}T)^{-1}$, and the normalization factor is given by the partition function (see Appendix B for mathematical details)
\begin{equation}
Z(B,\beta)=\sum_{\mathbf{n}} e^{ -\beta E_{\mathbf{n}} } = Z^+Z^-.
\label{eq37}
\end{equation}
Here, we have defined
\begin{equation}
Z^{\pm}=\frac{1}{2\sinh(\frac{\hbar\beta \omega_{\pm}}{2})},
\label{eq38}
\end{equation}
with
\begin{equation}
\omega_{\pm}=\Omega \pm \frac{\omega_B}{2}.
\label{eq39}
\end{equation}

From a similar analysis as in the previous section, we conclude that the heat absorbed by the system
from
the thermal reservoir is given by
\begin{eqnarray}
Q_{1\rightarrow 2} &=& \int_{B_{1}}^{B_{2}}\sum_{\mathbf{n}}E_{\mathbf{n}}(B)\frac{dp_{\mathbf{n}}(B,\beta_{H})}{dB}dB
\nonumber\\
&=& -\frac{\partial\ln\left(\frac{Z(B_{2},\beta_{H})}{Z(B_{1},\beta_{H})} \right)}{\partial\beta_{H}}
+ \beta_{H}^{-1}\ln\left(\frac{Z(B_{2},\beta_{H})}{Z(B_{1},\beta_{H})} \right)\nonumber\\
&=& E(B_2,\beta_H) - E(B_1,\beta_H) + \beta_{H}^{-1}ln\left[\frac{Z(B_2,\beta_H)}{Z(B_1,\beta_H)} \right]\nonumber\\
\label{eq40}
\end{eqnarray}
In the second line, we have done integration by parts, and we made direct use of the definition Eq.(\ref{eq37})
of the partition function. The final result follows from substituting the explicit expression for
the partition function Eq.(\ref{eq37}), and the definition of the ensemble-averaged energy
of the single-particle system $E = \langle \hat{H} \rangle = -\partial ln Z/\partial\beta$,
\begin{eqnarray}
E(B,\beta) = \frac{\hbar\omega_{+}}{2}coth\left(\frac{\beta\hbar\omega_{+}}{2}\right) +
\frac{\hbar\omega_{-}}{2}coth\left(\frac{\beta\hbar\omega_{-}}{2}\right)
\label{eq41}
\end{eqnarray}

Similarly, during the third stage of the cycle (see Fig.\ref{fig5}), the system is again brought into contact with a thermal
reservoir, but at a lower temperature $T_{C} < T_{H}$. Therefore, the probability distribution
of states in the ensemble is $p_{\mathbf{n}}(B,\beta_{C})$, as defined in Eq.(\ref{eq36}),
but with $T_{C}$ instead of $T_{H}$. The heat
released to the reservoir during this stage is given by the expression
\begin{eqnarray}
Q_{3\rightarrow 4}
= E(B_4,\beta_C) - E(B_3,\beta_C) + \beta_{C}^{-1}ln\left[\frac{Z(B_4,\beta_C)}{Z(B_3,\beta_C)}\right].\nonumber\\
\label{eq42}
\end{eqnarray}

\begin{figure}[tbp]
\centering
\epsfig{file=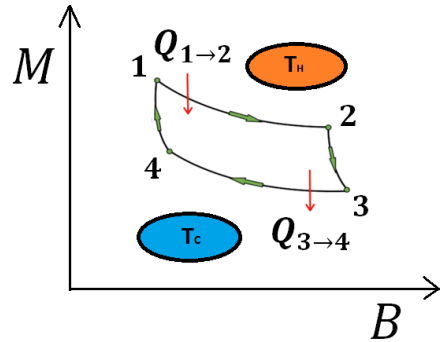,width=0.7\columnwidth,clip=}
\caption{(Color online) The magnetization of the system changes as a function of the applied external
magnetic field, along two isothermal and two iso-entropic trajectories of the cycle. The isothermal
trajectories are achieved by bringing the system into contact with
macroscopic thermal reservoirs at temperatures $T_{H}> T_{C}$, respectively.}
\label{fig6}
\end{figure}

The second and fourth stages of the cycle constitute iso-entropic trajectories (see Fig.~\ref{fig5}).
When substituting the Boltzmann distribution $p_{\mathbf{n}}(\beta,B)=\left[Z(\beta,B)\right]^{-1}\exp(-\beta E_{\mathbf{n}}(B))$ into the expression for the von Neumann entropy Eq.(\ref{eq11}), we obtain the relation
\begin{eqnarray}
S/k_{B} = \beta E + \ln Z(\beta,B).
\label{eq43}
\end{eqnarray}
Here, $E=\langle \hat{H} \rangle$ is the ensemble-average energy, as defined by Eq.(\ref{eq12}). The equation of state
is obtained from Eq.(\ref{eq43}) as
\begin{eqnarray}
M &=& -\left(\frac{\partial E}{\partial B}\right)_{S} = \beta^{-1}\frac{\partial}{\partial B}\ln Z(\beta,B)\nonumber\\
&=& -\mu_{B}\frac{\omega_{+}}{\Omega}coth\left(\frac{\beta\hbar\omega_{+}}{2} \right)+\mu_{B}\frac{\omega_{-}}{\Omega}coth\left(\frac{\beta\hbar\omega_{-}}{2} \right),\nonumber\\
\label{eq44}
\end{eqnarray}
with $M$ the ensemble-average magnetization as a function of the
external magnetic field (see Fig.~\ref{fig6}), and $\mu_{B}=e\hbar/(2m^{*})$ the Bohr magneton.
In the last line, we made use of the explicit analytical expression Eq.(\ref{eq37}) for the partition function to calculate
the derivative.
The work performed during the second stage of the process is
\begin{eqnarray}
W_{2\rightarrow 3} = E(B_{2},\beta_{H}) - E(B_{3},\beta_{C})
\label{eq45}
\end{eqnarray}
We are now in conditions to discuss the second and fourth stages of the Carnot cycle. These iso-entropic trajectories impose
implicit conditions for the intensities of the magnetic field,
\begin{eqnarray}
\Delta S_{2\rightarrow 3} &=& S(B_{3},\beta_{C}) - S(B_{2},\beta_{H})\nonumber\\
\Delta S_{4\rightarrow 1} &=& S(B_{1},\beta_{H}) - S(B_{4},\beta_{C})
\label{eq46}
\end{eqnarray}
Combining Eqs.(\ref{eq46}), and further expressing the entropies in terms of Eq.(\ref{eq43}), we obtain the condition
\begin{eqnarray}
&&\beta_{C}\left[E(B_{3},\beta_{C})-E(B_{4},\beta_{C})\right] + ln\left[\frac{Z(B_{3},\beta_{C})}{Z(B_{4},\beta_{C})}\right]\nonumber\\
&&= \beta_{H}\left[E(B_{2},\beta_{H})-E(B_{1},\beta_{H})\right] + ln\left[\frac{Z(B_{2},\beta_{H})}{Z(B_{1},\beta_{H})}\right].\nonumber\\
\label{eq47}
\end{eqnarray}

The fourth and final stage of the cycle also corresponds to an iso-entropic trajectory (see Fig.~\ref{fig5}) where $l_{B_{4}}\rightarrow l_{B_{1}}$, and the work performed
by the system is given by
\begin{eqnarray}
W_{4\rightarrow 1} &=& E(B_{4},\beta_{C}) - E(B_{1},\beta_{H}).
\label{eq48}
\end{eqnarray}

The efficiency of the quantum Carnot cycle is given by
\begin{eqnarray}
\eta^{C} &=& 1 - \frac{Q_{3\rightarrow 4}}{Q_{1\rightarrow 2}} = 1 - \frac{T_{C}}{T_{H}},
\label{eq49}
\end{eqnarray}
where we have made use of Eq.(\ref{eq40}) and Eq.(\ref{eq42}) to obtain the second line.
Remarkably, the efficiency is identical to the classical Carnot cycle. This result
is in agreement with what we found in a recent work, where the efficiency for
a mechanically driven quantum heat engine based on a relativistic Dirac particle was studied \cite{Munoz_Pena_012}.

\section{Conclusions}

In this work, we explored the possibility of constructing a single-particle quantum heat engine,
by means of combining the confining effects of a cylindrical potential well and an externally
imposed magnetic field. The overall confinement length scale is characterized by the Landau radius $l_{e,B}$, which can be quasi-statically modified by tuning the intensity of the external magnetic field $B$.
We considered two different schemes to implement this idea: The Iso-energetic cycle and the quantum Carnot cycle. For both cases, we obtained explicit analytical expressions for the quantum engine efficiencies. In particular, for the Iso-energetic cycle, we showed that in the high magnetic field regime
the efficiency achieves the asymptotic limit $\eta \sim 1 - 1/\alpha^2$, in agreement with
results previously reported in the literature for mechanically driven quantum engines \cite{Bender_02},
even in the low-energy regime for relativistic Dirac particles \cite{Munoz_Pena_012}.
The generality of this expression
suggests that it may represent a universal upper theoretical limit for the efficiency
of any quantum heat engine operating under the Iso-energetic cycle scheme. The formal proof of this conjecture, however, goes beyond the scope of the present work.

For the quantum Carnot cycle, we proved that its efficiency depends only on the ratio between the temperatures of the cold and hot reservoirs $\eta^C = 1 - T_C/T_H$, exactly as in the classical
Carnot cycle for macroscopic heat engines. This remarkable result was also obtained
in the relativistic limit for a mechanically driven quantum heat engine \cite{Munoz_Pena_012},
and hence its apparent universal character reflects the conceptual robustness of Thermodynamics.
On the other hand, if the reservoirs themselves are considered as quantum mechanical objects,
these can in principle be prepared into quantum coherent non-thermal states\cite{Scully_011,Scully_03},
or into squeezed thermal
states\cite{Lutz_014}. Under such conditions, theoretical studies show that the aforementioned Carnot limit for
the efficiency can be surpassed\cite{Lutz_014,Scully_011,Scully_03}.

This work is mainly intended as a proof of concept rather than as a practical
implementation protocol. However, we believe that experimentally it would be
easier to control the intensity of an applied magnetic field, than to impose
a mechanical force upon a nanosystem. For this reason, we have focused the modeling aspects in the context of a GaAs quantum dot as a prototype
system for a cylindrical potential well, on the hope that the idea proposed here
may be attractive for experimental scientists interested in a future practical implementation
of this concept.

\section*{Acknowledgements}

E.M. acknowledges financial support from Fondecyt Grant 1141146. F.J.P.
acknowledges financial support from a Conicyt fellowship.

\section*{Appendix A}

In this appendix, we present the detailed derivation of Eq.(\ref{eq22}).
For a two-level system experiencing an iso-energetic process
that connects states $1 \rightarrow 2$, the probabilities satisfy
\begin{eqnarray}
p_{\mathbf{1}}(B) = \frac{E_{\mathbf{2}}(B_{1}) - E_{\mathbf{2}}(B)}{E_{\mathbf{1}}(B) - E_{\mathbf{2}}(B)} + \frac{E_{\mathbf{1}}(B_{1}) - E_{\mathbf{2}}(B_{1})}
{E_{\mathbf{1}}(B) - E_{\mathbf{2}}(B)}p_{\mathbf{1}}(B_{1}),\nonumber\\
\label{eq50}
\end{eqnarray}
with $p_{\mathbf{2}}(B) = 1 - p_{\mathbf{1}}(B)$. The expression for the heat exchanged during the process is, from Eq.(\ref{eq20})
\begin{eqnarray}
Q_{1\rightarrow 2} &=& \int_{B_{1}}^{B_{2}}\left(p_{1}(B)\frac{d E_{1}}{d B} +
p_{2}(B)\frac{d E_{2}}{d B} \right)dB\nonumber\\
&=& \int_{B_{1}}^{B_{2}}\left(p_{1}(B)\frac{d (E_{1} - E_{2})}{d B}
+ \frac{d E_{2}}{d B} \right)dB.\nonumber\\
\label{eq51}
\end{eqnarray}
Upon substitution of Eq.(\ref{eq50}) into Eq.(\ref{eq51}), after some algebra one obtains
\begin{eqnarray}
&&Q_{1\rightarrow 2} = \left[E_{\mathbf{2}}(B_{1}) + \left(E_{\mathbf{1}}(B_{1})
- E_{\mathbf{2}}(B_{1}) \right)p_{\mathbf{1}}(B_{1}) \right]\nonumber\\
&&\times\int_{B_{1}}^{B_{2}}\frac{\frac{d\left(E_{\mathbf{1}} - E_{\mathbf{2}}\right)}{dB}}
{E_{\mathbf{1}}(B) - E_{\mathbf{2}}(B)}
+\int_{B_{1}}^{B_{2}}\frac{E_{\mathbf{1}}\frac{dE_{\mathbf{2}}}{dB}- E_{\mathbf{2}}
\frac{dE_{\mathbf{1}}}{dB}}{E_{\mathbf{1}}(B) - E_{\mathbf{2}}(B)}dB.\nonumber\\
\label{eq52}
\end{eqnarray}
The first integral is elementary, and we obtain the expression
\begin{eqnarray}
&&Q_{1\rightarrow 2} = \left[E_{\mathbf{2}}(B_{1}) + \left(E_{\mathbf{1}}(B_{1})
- E_{\mathbf{2}}(B_{1}) \right)p_{\mathbf{1}}(B_{1}) \right]
\nonumber\\
&&\times ln\left[\frac{E_{\mathbf{1}}(B_{2}) - E_{\mathbf{2}}(B_{2})}
{E_{\mathbf{1}}(B_{1}) - E_{\mathbf{2}}(B_{1})} \right]
+\int_{B_{1}}^{B_{2}}\frac{E_{\mathbf{1}}\frac{dE_{\mathbf{2}}}{dB}
- E_{\mathbf{2}}\frac{dE_{\mathbf{1}}}{dB}}{E_{\mathbf{1}}(B)
- E_{\mathbf{2}}(B)}dB.\nonumber\\
\label{eq53}
\end{eqnarray}
In particular, when the azimuthal quantum numbers of both states are $m_{1}=m_{2}=0$, then the second integral in Eq.(\ref{eq53}) trivially vanishes and one has
\begin{eqnarray}
&&Q_{1\rightarrow 2} = \left[E_{2}(B_{1}) + \left(E_{1}(B_{1}) - E_{2}(B_{1}) \right)p_{1}(B_{1}) \right]
\nonumber\\
&&\times ln\left[\frac{E_{1}(B_{2}) - E_{2}(B_{2})}{E_{1}(B_{1}) - E_{2}(B_{1})} \right]\,({\rm{if}}\,\,m_1 = m_2 =0).
\label{eq54}
\end{eqnarray}

\section*{Appendix B}

In this appendix, we present the mathematical details to obtain the partition function for the ensemble
of single-particle systems under consideration.
The spectrum for the system, corresponding to effective Landau levels, is determined by radial $n_{\rho}=0,1,\ldots$
and azimuthal $m=0,\pm 1,\pm 2,\ldots$ quantum numbers,
\begin{eqnarray}
E_{n_{\rho},m}(B) = \hbar\Omega\left(2 n_{\rho}+|m|+1 \right) - \frac{m}{2}\hbar\omega_{B}.
\label{eq54}
\end{eqnarray}
It is convenient to re-parameterize this quantum numbers in terms of a pair of integers $n_{+}$, $n_{-}$, defined
as
\begin{equation}
n_+=\frac{1}{2}(2n_{\rho}+\mid{m}\mid-m),\,\,\,\,
n_-=\frac{1}{2}(2n_{\rho}+\mid{m}\mid+m)
\label{eq55}
\end{equation}
with $n_{\pm}=0,1,\ldots,\infty$.
Combining both definitions, we obtain
\begin{equation}
m=n_--n_+ \qquad n_{\rho}=\frac{1}{2}(n_+ + n_-).
\label{eq56}
\end{equation}
The energy spectrum can be rewritten as:
\begin{eqnarray}
E_{n_{\rho},m}(B) = E(n_+)+E(n_-),
\label{eq57}
\end{eqnarray}
where we defined
\begin{equation}
E_{n_\pm} = \hbar \omega_{\pm}(n_{\pm} \pm \frac{1}{2}).
\label{eq58}
\end{equation}
Here,
\begin{eqnarray}
\omega_{\pm} = \Omega \pm \frac{\omega_{B}}{2}.
\label{eq59}
\end{eqnarray}
The partition function is then calculated as
\begin{eqnarray}
Z(B,\beta) &=& \sum_{n_{\rho},m}e^{-\beta E_{n_{\rho,m}}(B)}\nonumber\\
&=& \sum_{n_{+}=0}^{\infty}\sum_{n_{-}=0}^{\infty}e^{-\beta\left(E_{n_+}+E_{n_-} \right)}
= Z^{+}Z^{-}.
\label{eq60}
\end{eqnarray}
Here, we have defined
\begin{eqnarray}
Z^{\pm} = \frac{1}{2sinh\left(\frac{\beta\hbar\omega_{\pm}}{2} \right)}.
\label{eq61}
\end{eqnarray}

\bibliographystyle{apsrev}

\begin{thebibliography}{24}
\bibitem[{not()}]{note1}
\bibinfo{note}{A necessary condition for entropy to remain constant is
  $\sum_{n=1}^{\infty}(1 + \ln(p_{n}(L))dp_{n}/dL=0$. This is clearly less
  stringent than the sufficient condition $dp_{n}/dL = 0$ for all n.}
\end{thebibliography}

\end{document}